\documentclass[journal=nalefd,manuscript=letter]{achemso}

\usepackage{achemso}

\usepackage{color}
\usepackage{float}
\usepackage{array}
\usepackage{lmodern}
\makeatletter
\@ifundefined{textcolor}{}
{%
 \definecolor{BLACK}{gray}{0}
 \definecolor{WHITE}{gray}{1}
 \definecolor{RED}{rgb}{1,0,0}
 \definecolor{GREEN}{rgb}{0,1,0}
 \definecolor{BLUE}{rgb}{0,0,1}
 \definecolor{CYAN}{cmyk}{1,0,0,0}
 \definecolor{MAGENTA}{cmyk}{0,1,0,0}
 \definecolor{YELLOW}{cmyk}{0,0,1,0}
}

\setkeys{acs}{usetitle=true}
\setkeys{acs}{keywords=true}
\setkeys{acs}{super=true}
\setkeys{acs}{articletitle=false}

\usepackage[version=3]{mhchem} 
\usepackage[T1]{fontenc}       

\author{Natsumi~Komatsu}%
\affiliation{Department of Electrical and Computer Engineering, Rice University, Houston, Texas 77005, USA}

\author{Motonori~Nakamura}
\affiliation{Department of Systems, Control and Information Engineering, National Institute of Technology, Asahikawa College, Asahikawa, Hokkaido 071-8142, Japan}

\author{Saunab~Ghosh}%
\affiliation{Department of Electrical and Computer Engineering, Rice University, Houston, Texas 77005, USA}

\author{Daeun~Kim}%
\affiliation{Department of Electronics for Informatics, Hokkaido University, Sapporo, Hokkaido 060-0814, Japan}%

\author{Haoze~Chen}%
\affiliation{Department of Electrical and Computer Engineering, Rice University, Houston, Texas 77005, USA}

\author{Atsuhiro Katagiri}
\affiliation{Department of Physics, Tokyo Metropolitan University, Tokyo 192-0372, Japan}%

\author{Yohei~Yomogida}%
\affiliation{Department of Physics, Tokyo Metropolitan University, Tokyo 192-0372, Japan}%

\author{Weilu~Gao}%
\affiliation{Department of Electrical and Computer Engineering, Rice University, Houston, Texas 77005, USA}

\author{\\Kazuhiro~Yanagi}%
\affiliation{Department of Physics, Tokyo Metropolitan University, Tokyo 192-0372, Japan}%

\author{Junichiro Kono}%
\email{kono@rice.edu}
\affiliation{Department of Electrical and Computer Engineering, Rice University, Houston, Texas 77005, USA}
\alsoaffiliation{Department of Physics and Astronomy, Rice University, Houston, Texas 77005, USA}
\alsoaffiliation{Department of Materials Science and NanoEngineering, Rice University, Houston, Texas 77005, USA}

\title[An \textsf{achemso} demo]
  {Groove-Assisted Global Spontaneous Alignment of Carbon Nanotubes in Vacuum Filtration}


\begin{document}

\setlength{\fboxrule}{0 pt}








\pagebreak

\begin{abstract}
Ever since the discovery of carbon nanotubes (CNTs), it has long been a challenging goal to create macroscopically ordered assemblies, or crystals, of CNTs that preserve the one-dimensional quantum properties of individual CNTs on a macroscopic scale. Recently, a simple and well-controlled method was reported for producing wafer-scale crystalline films of highly aligned and densely packed CNTs through spontaneous global alignment that occurs during vacuum filtration [\textit{Nat.\ Nanotechnol}.\ \textbf{11}, 633 (2016)]. However, a full understanding of the mechanism of such global alignment has not been achieved. Here, we report results of a series of systematic experiments that demonstrate that the CNT alignment direction can be controlled by the surface morphology of the filter membrane used in the vacuum filtration process.  More specifically, we found that the direction of parallel grooves pre-existing on the surface of the filter membrane dictates the direction of the resulting CNT alignment.  Furthermore, we intentionally imprinted periodically spaced parallel grooves on a filter membranes using a diffraction grating, which successfully defined the direction of the global alignment of CNTs in a precise and reproducible manner.
\end{abstract}

\pagebreak

\section{Main Text}

{\label{874460}}

The one-dimensional (1D) quantum confinement of electrons and phonons in single-wall carbon nanotubes (SWCNTs) leads to extremely anisotropic electronic, mechanical, thermal, magnetic, and optical properties, which can be useful for a variety of device applications.\cite{DresselhausetAl01Book,JorioetAl08Book}  However, such 1D properties are often lost in macroscopic assemblies of SWCNTs, due to imperfect alignment and low packing fractions.  Recently, He, Gao, and coworkers have developed a method, known as controlled vacuum filtration (CVF), for producing wafer-scale (centimeter-sized) crystalline films of SWCNTs in which nanotubes are nearly perfectly aligned (with nematic order parameter $S \sim 1$) and maximally packed ($\sim$1 nanotube per cross-sectional area of 1\,nm$^2$).\cite{HeetAl16NN,KomatsuetAl17AFM,GaoKono19RSOS}  The CVF method has been subsequently used by other groups to obtain similar aligned CNT films.\cite{ChiuetAl17NL,FukuharaetAl18APL,HoetAl18PNAS,RobertsetAl19NL,ChenetAl19ACS-AMI,IchinoseetAl19NL,WalkeretAl19NL}  However, the mechanism that produces such global spontaneous alignment during vacuum filtration is not fully understood.  In particular, what determines the direction of CNT alignment has been a matter of controversy.

He \textit{et al}.\ observed that a hydrophilic coating layer on the filter membrane was crucial for achieving global alignment.  Hydrophilic coating makes the membrane surface negatively charged, and, as a result, negatively charged CNTs in the aqueous suspension should be electrostatically repelled from the surface. At the same time, the gravity as well as the van der Waals attraction from the surface should push the CNTs downward. There should thus be a certain vertical location where the potential energy is minimum, which is horizontally extended throughout the membrane surface globally.  CNTs would be trapped there, floating from the surface and freely rotating in the horizontal direction.  

Based on these considerations, He \textit{et al}.\ proposed a model, which we refer to as the Global ``Plate Tectonics'' (GPT) model in this article.  As the filtration process proceeds, the CNT density increases and the chance that two CNTs meet increases.  When two CNTs meet, they rotate to align with each other to form a bundle.  When two bundles meet, they rotate to align with each other to form a larger bundle.  Therefore, as time progresses, larger and larger bundles, or plate-like ``islands,'' appear at different locations on the filter membrane.  CNTs are aligned in one direction within each island, but the alignment directions of different islands are random. These islands are still floating and freely rotating in the horizontal direction, and hence, when two islands meet, they rotate to align with each other to form a larger island, or a ``continent.''  Different continents meet, rotate, and merge, eventually forming a wafer-scale monodomain film in which all CNTs are pointing in the same direction.


In this article, we first present experimental results that cannot be explained by the GPT model.  Specifically, we cut the whole membrane into four pieces before filtration and still found global monodomain alignment, i.e., the CNT alignment directions on the four separate pieces were still the same.  This finding suggests that the alignment direction is determined by some characteristic of the entire filter membrane.  Indeed, we found that the alignment direction was predetermined by the direction of a large number of parallel grooves that exist on most commercial filter membranes, as pointed out by Chi \textit{et al}.\cite{ChiuetAl17NL}  Although He \textit{et al}.\ were also aware of such grooves on their membranes, they did not observe clear correlation between the groove direction and the CNT alignment direction.\cite{HeetAl16NN}  We characterized the dimensions and distributions of grooves using atomic force microscopy (AFM) and found that the groove direction dictates the direction of CNT alignment when grooves exist.  When we eliminated grooves through heating, no global alignment was observed although there was some local alignment; multiple domains appeared, and different domains had different alignment directions, similar to an earlier report.\cite{DanetAl12IECR}  Finally, we intentionally created grooves on the filter membrane by pressing a diffraction grating against it after the heating treatment.  The direction of resulting CNT alignment in this case was the same as the direction of the intentionally created grooves.  This procedure, groove-assisted CVF, provides a simple and reproducible way to align CNTs in a desired direction.



We cut a filter membrane into four pieces (Pieces 1-4), as schematically shown in Figure~1a, so that there can be no communication between the four pieces. 
If global CNT alignment occurs in a way consistent with the GPT model, the CNT alignment direction on each piece must be independent.  Namely, we should expect each piece to have monodomain CNT alignment, but the alignment directions of the four pieces should be independent of each other.  Figure~1b shows the polarization dependence of light absorbance for the CNT films created on Pieces 1-4.  See the Experimental Section for characterization details.  Absorbance should have a maximum (minimum) value when the incident light is polarized parallel (perpendicular) to the CNT alignment direction.  The data in Figure~1b unequivocally demonstrates that the four pieces have the same alignment direction, in disagreement with what one expects from the GPT model.

\begin{figure}[h!]
\begin{center}
\includegraphics[scale=0.5]{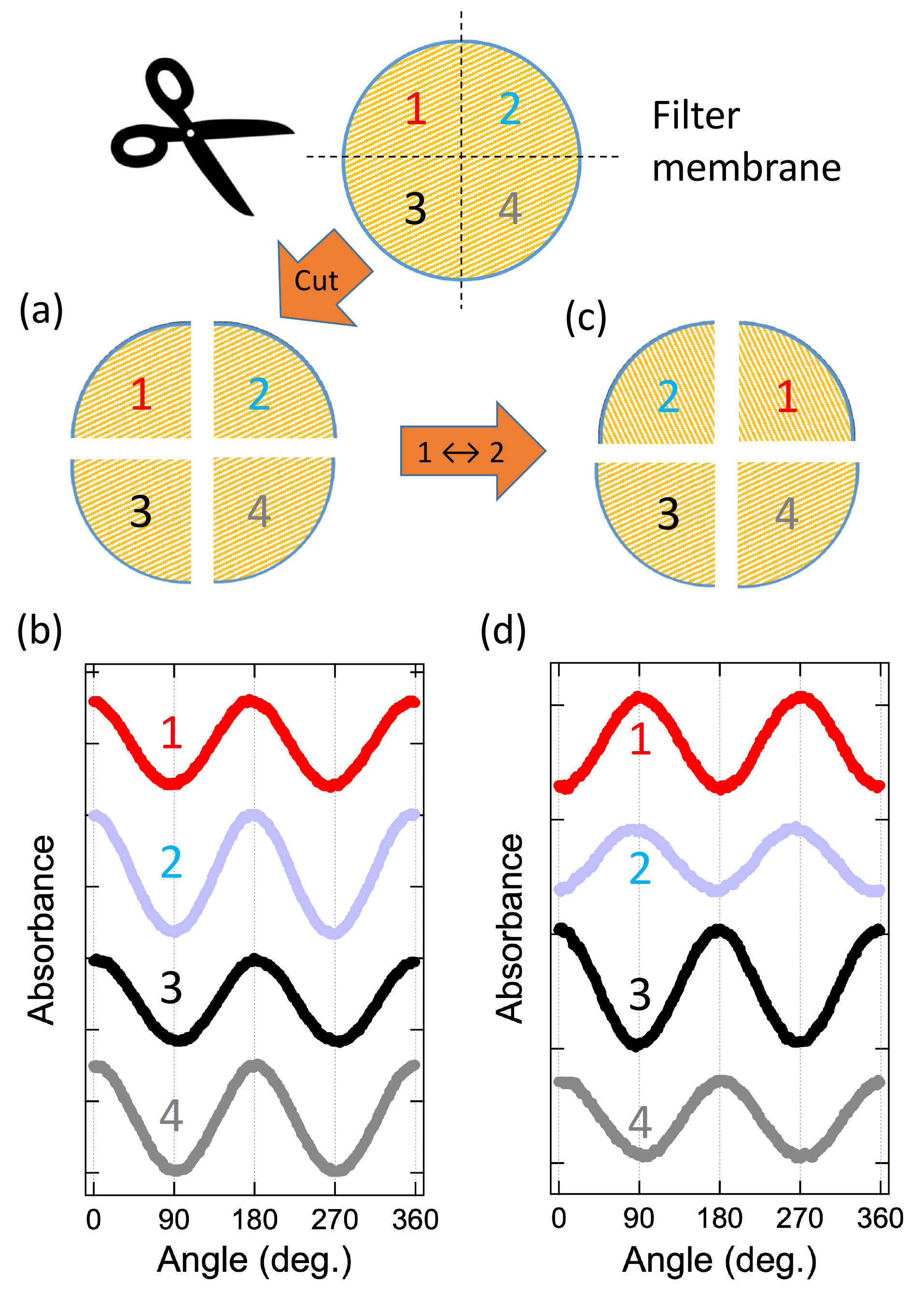}
\caption{Failure of the GPT model. (a)~A filter membrane was cut into four pieces.  (b)~Polarization angle dependent absorbance for a film made from the cut filter membrane shown in (a). The numbers given to the traces correspond to the piece numbers in (a).  Note that we did not rotate the piece but scanned each piece while maintaining the initial polarization direction along the $y$-axis of the image.  The traces are vertically offset for clarity.  There is no phase shift among the four traces, meaning that the CNT alignment directions of the four pieces are the same even though they were physically separated from each other before filtration.  (c)~Pieces 1 and 2 were interchanged.  Note that this required these pieces to be rotated by 90 degrees.  (d)~Polarization angle dependent absorbance for a film made from the cut filter membrane shown in (c).  The numbers given to the traces correspond to the piece numbers in (c).  The traces are vertically offset for clarity.  The data indicates that the CNT alignment direction of Pieces 1 and 2 is the same and perpendicular to that of Pieces 3 and 4.  The data in (b) and (d) are inconsistent with the GPT model, suggesting instead that the alignment direction is determined by some local characteristic of the filter membrane.}
\label{Fig1}%
\end{center}
\end{figure}

We conducted another experiment to further test the GPT model.  After cutting a membrane into four pieces, we exchanged the top two pieces (Pieces 1 and 2), which required 90 degree rotation of each piece, as schematically shown in Figure~1c.  
Again, if the GPT mechanism is at work, the CNT alignment directions of the four pieces should be totally uncorrelated with each other.  However, Figure~1d shows that the four directions are strongly correlated.  Specifically, the top two (Pieces 1 and 2) have the same alignment direction, the bottom two (Pieces 3 and 4) have the same alignment direction, and there is a 90-degree shift in alignment direction between the top and bottom pieces.  This result, together with the result shown in Figure~1b, invalidates the GPT model, demonstrating instead that the alignment direction is determined by some local characteristic of the membrane.

\newpage

As mentioned earlier, there are always a number of parallel-oriented grooves found on commercial filter membranes, which were presumably created during the manufacturing process.  Figure~2a shows an optical microscopy image, while Figures~2b and 2c show atomic force microscopy (AFM) images, of the surface of a typical filter membrane.  There are ``macrogrooves'' and ``microgrooves.''  A macrogroove consists of many microgrooves.  In Figure~2a, white arrows indicate macrogrooves, which are roughly periodically located.  The width of each macrogroove varied from a few $\mu$m (in which case there were only a few microgrooves) to 20~$\mu$m, and the distance between the centers of two adjacent macrogrooves (the intermacrogroove distance) varied from 30~$\mu$m to 90~$\mu$m. 
Figure~2b shows a macrogroove consisting of several microgrooves. Microgrooves were not continuous across the filter membrane, some of them disappearing in the middle.  We used large-magnification AFM images, such as Figure~2c, to estimate the depths and widths of microgrooves; see Supplementary Information for more details. The depth of microgrooves varied from a few nm to 60~nm (the average being 13~nm), and the width varied from 100~nm to 700~nm (the average being 290~nm).

\begin{figure}[h!]
\begin{center}
\includegraphics[scale=1.00]{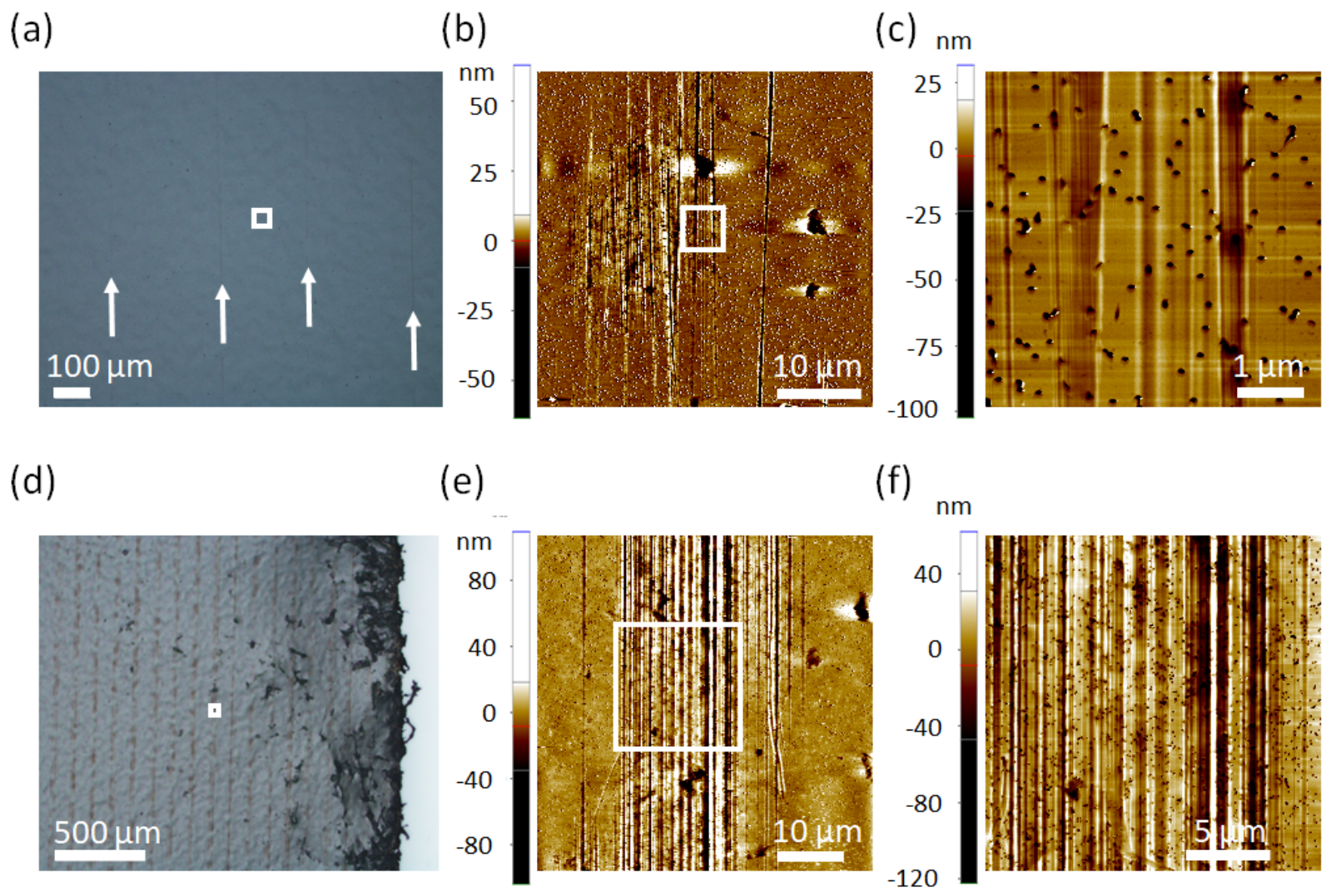}
\caption{Macrogrooves and microgrooves on filter membranes.  (a)~Optical microscopy image of a filter membrane with a typical groove density.  The white arrows indicate macrogrooves.  (b)~An AFM image of the area indicated by the white square in (a), showing a macrogroove containing several microgrooves. (c)~A magnified AFM image of the area indicated by the white square in (b), showing microgrooves. (d)~Optical microscopy image of a filter membrane with an unusually high groove density. (e)~An AFM image of the area indicated by a white square in (d), showing a macrogroove containing microgrooves. (f)~A magnified AFM image of the area indicated by the white square in (e), showing microgrooves.}
\label{Fig2}%
\end{center}
\end{figure}


The density of grooves varied from batch to batch.  Figures~2d--2f show images of a filter membrane with an unusually high groove density.  As Figure~2d shows, the macrogrooves on this membrane are more periodic than that on the typical membrane shown in Figure~2a. Figure~2e shows that the number of microgrooves within one macrogroove is high in this membrane; a macrogroove consists of twenty to thirty microgrooves.  The width of a macrogroove was $\sim$30~$\mu$m, and the intermacrogroove distance was $\sim$90~$\mu$m.  We used a higher magnification AFM image (Figure~2f) to estimate the depths and widths of microgrooves (see Supplementary Information for details).  The depth of a microgroove varied from a few nm to 70~nm (the average being 24~nm), and the width varied from 200~nm to 900~nm (the average being 503~nm).

Figure~3a shows an AFM image of an aligned CNT film on a filter membrane.  The filter membrane had a high density of grooves (from the same batch as that show in Figures~2d, 2e, and 2f).  Figures~3b and 3c show magnified images of two different areas of the membrane.  
Figure~3b shows an area that contains microgrooves, while Figure~3c shows an area that do not contain grooves; in both Figures~3b and 3c, the direction of grooves is along the $y$-axis of the images.  The white arrows indicate the CNT alignment direction.  These images show that the CNT alignment direction is the same as the groove direction in both areas.

\begin{figure}[h!]
\begin{center}
\includegraphics[scale=0.86]{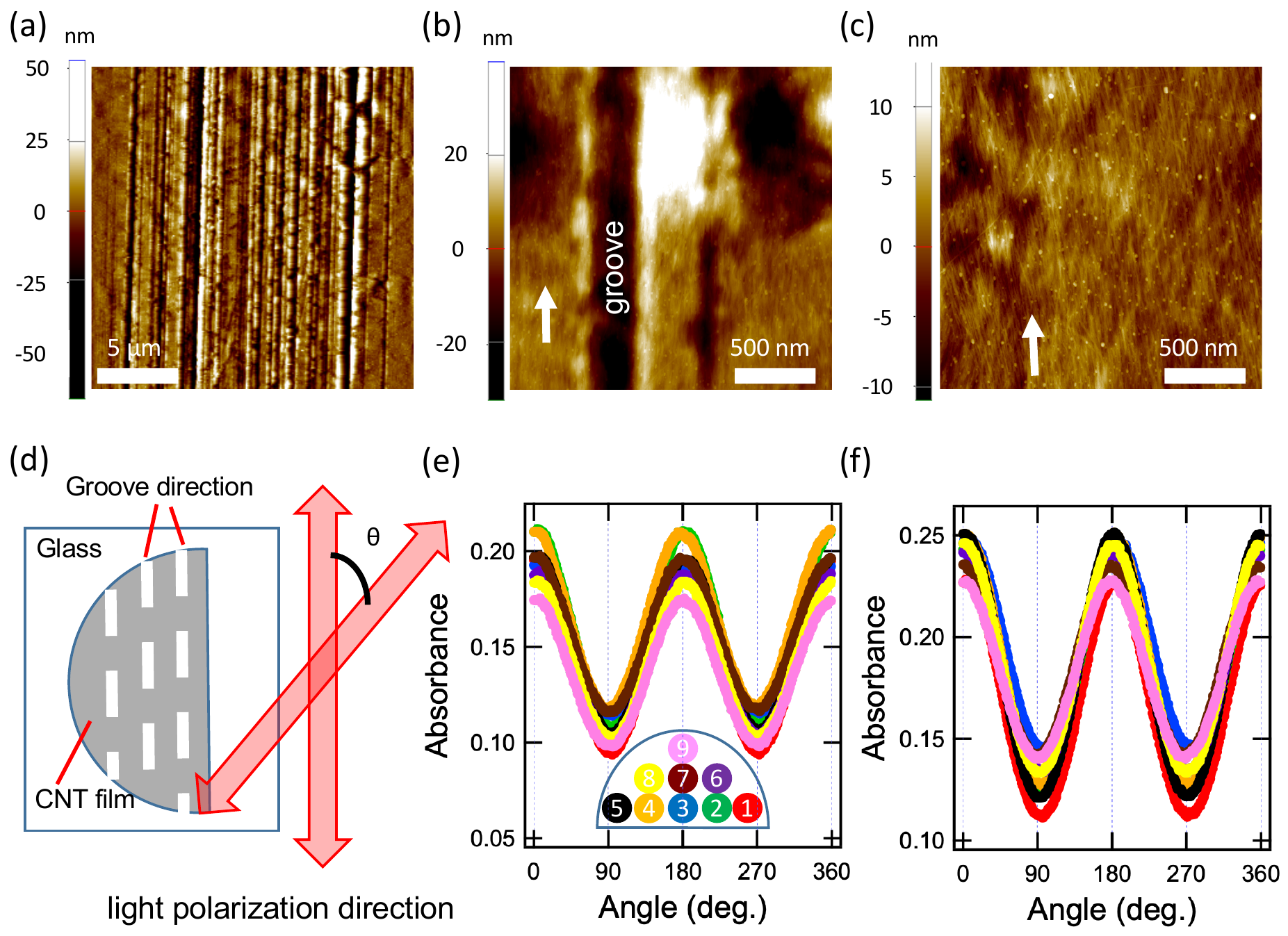}
\caption{Agreement of the CNT alignment direction and the groove direction. (a)~An AFM image of an aligned CNT film made on a filter membrane with a high groove density. 
Magnified AFM images of the film, still on the membrane, taken in an area (b)~with grooves and (c)~without grooves.  The groove direction was along the $y$-axis of the images in (b) and (c). 
The white arrows indicate the CNT alignment direction. 
(d)~Schematic showing the configuration of the polarization-dependent absorbance measurements used to compare the groove and CNT alignment directions.  The incident light polarization and the groove direction match at $\theta = 0^\circ$.  (e) and (f)~Polarization angle dependent absorbance for aligned SWCNT films made on (e) a typical filter membrane  and (f) a filter membrane with high groove density.  Each trace in (e) and (f) corresponds to a different spot on the SWCNT film schematically indicated by a circle of the corresponding color in the inset of (e).  Maximum absorbance occurs at $\theta = 0^\circ$ and $180^\circ$ in (e) and (f), proving that the CNT alignment direction is the same as the groove direction.}
\label{Fig3}%
\end{center}
\end{figure}


We then characterized the CNT alignment direction on a macroscopic scale with polarization-dependent absorbance measurements. We aligned the groove direction with the incident light polarization direction at angle $\theta$ = 0$^\circ$, as schematically shown in Figure~3d.  
Since absorbance is maximum (minimum) when the light polarization is parallel (perpendicular) to the CNT alignment direction, if the CNT alignment direction coincides with the groove direction, one should see maximum absorbance at $\theta$ = 0$^\circ$ and minimum absorbance at $\theta$ = 90$^\circ$. Figures~3e and 3f confirm this expected behavior for films made on filer membranes with low and high groove densities, respectively.  Each trace corresponds to a particular spot within the film, as indicated in the inset of Figure~3e, demonstrating that both the degree and direction of alignment are nearly constant across the whole film.  
Both the images in Figures~3b and 3c and the polarization-dependent absorbance data in Figures~3e and 3f unambiguously show that the groove direction dictates the CNT alignment direction.


To further demonstrate how crucial the existence of grooves is in CNT global alignment, we intentionally eliminated them before filtration.  We used the heating treatment process described in the Experimental Section to eliminate grooves.  Figure~4a shows an SEM image of a typical filter membrane without any treatment, exhibiting microgrooves, consistent with Figure~2b, whereas Figure~4b shows an SEM image of a membrane after heating treatment, displaying no trace of grooves.  Figures~4c and 4d show SEM images of CNT films made with the filter membranes with and without grooves, shown in Figures~4a and 4b, respectively.  Figure~4c reveals global alignment of CNTs over the entire area in the same direction as the groove direction, as shown in Figure~3.  However, no global alignment is found in Figure~4d.  Instead, the film shows multiple domains of aligned CNTs, similar to a previous report on slow vacuum filtration.\cite{DanetAl12IECR}  While CNTs within each domain are aligned in a common direction, different domains have different alignment directions, uncorrelated with the direction of the original groove direction before the heating treatment process.  

\begin{figure}[h!]
\begin{center}
\includegraphics[scale=0.86]{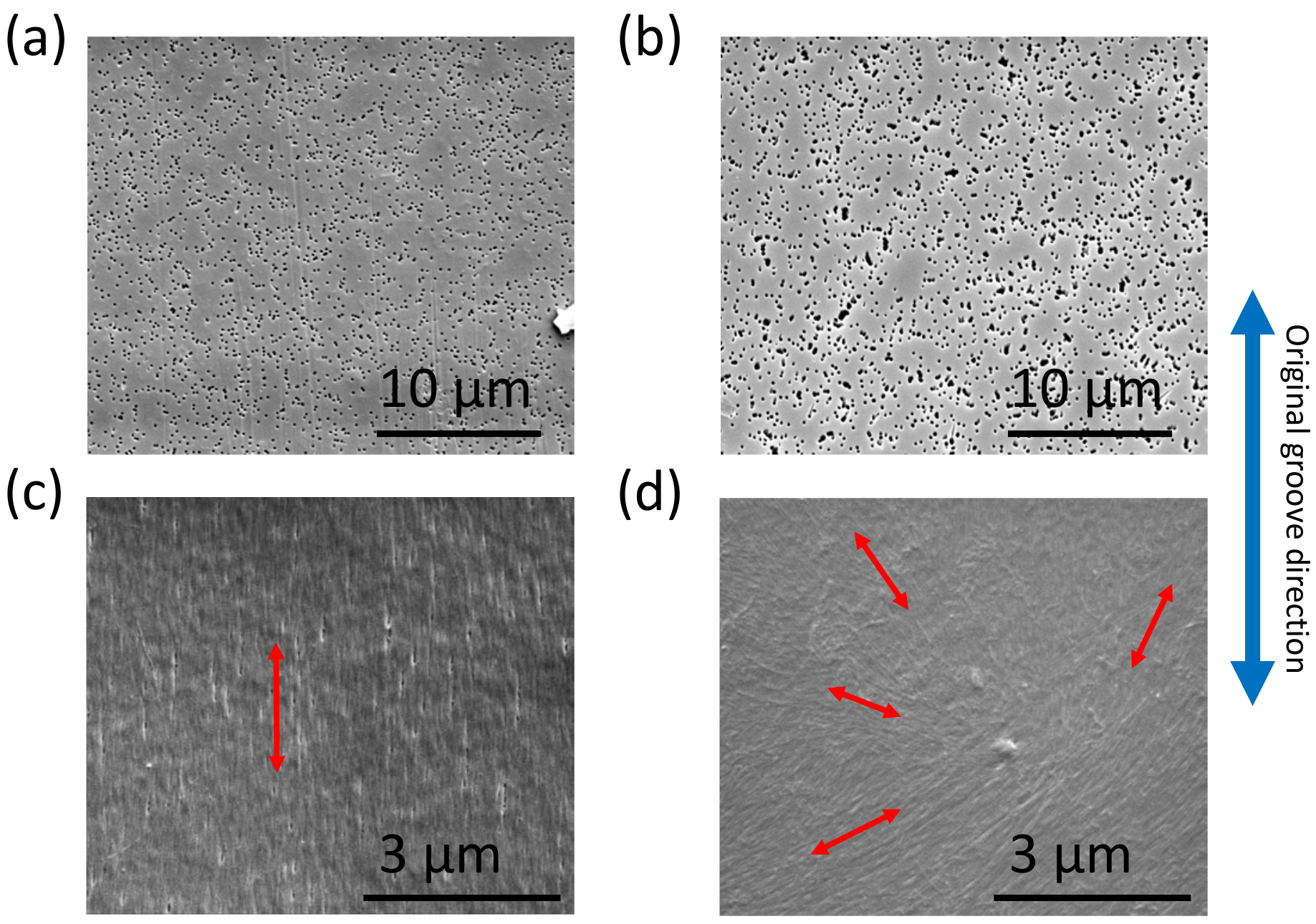}
\caption{The disappearance of global alignment upon elimination of grooves through heating.  The blue arrow indicates the direction of the original grooves before the heating treatment, applied to all four images. (a)~SEM image of a typical filter membrane without any treatment, showing microgrooves. (b)~SEM image of a filter membrane after heating treatment, exhibiting no grooves. (c)~SEM image of a CNT film made on the membrane shown in (a), demonstrating global alignment along the groove direction.  The red arrow indicates the CNT alignment direction, which is parallel to the blue arrow. (d)~SEM image of a CNT film made on the membrane shown in (b). There is no global alignment, but domains of aligned CNTs exist with random alignment directions.  The red arrows indicate the alignment directions of CNTs in the domains. }
\label{Fig4}%
\end{center}
\end{figure}



\begin{figure}[h!]
\begin{center}
\includegraphics[scale=0.83]{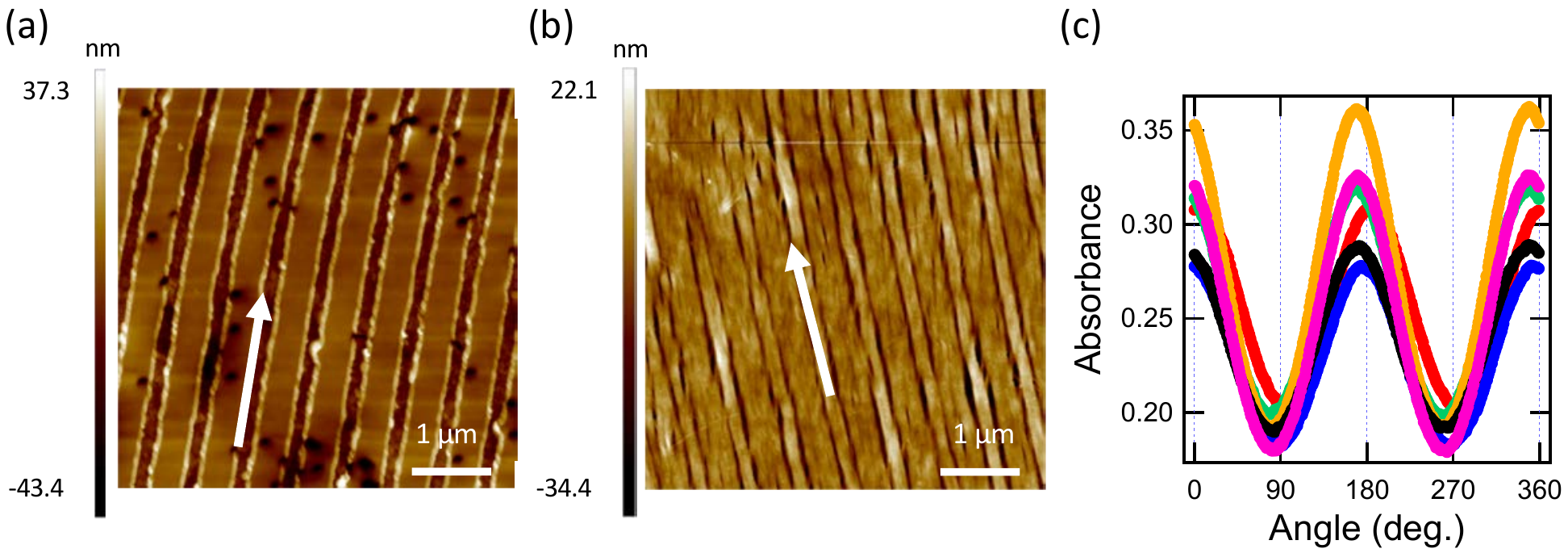}
\caption{Fabricating grooves on filter membranes. (a)~AFM image of a filter membrane with grooves intentionally created by a grating with a groove density of 1800/mm. The white arrow indicates the created groove direction. (b)~AFM image of a SWCNT film made on the filter membrane with intentionally created grooves. The white arrow indicates the groove direction. (c)~Polarization-dependent absorbance for the film shown in (b).  Each trace corresponds to a different spot on the film.  The absorbance becomes maximum at $\theta$ = 0$^\circ$ and minimum at $\theta$ = 90$^\circ$, demonstrating that the CNT alignment direction agrees with the intentionally created groove direction.}
\label{Fig5}%
\end{center}
\end{figure}

Furthermore, we intentionally created grooves on filter membranes after eliminating the pre-existing grooves by heating. Figure~5a shows an AFM image of a filter membrane with grooves made with a grating with a groove density of 1800/mm.  As demonstrated by this image, the groove pattern of the grating was successfully projected onto the filter membrane.  
Figure~5b shows an AFM image of a CNT film made on the filter membrane shown in Figure~5a.  The groove pattern on the filter membrane is visible even after being covered by the CNTs.  As shown by the white arrow in Figure~5b, the CNT alignment direction coincided with the fabricated groove direction in the entire area, drastically different from local domains observed in Figure~4d.  See Supplementary Information for results with gratings with different groove densities.  

We then characterized the CNT alignment direction on a macroscopic scale by polarization-dependent absorbance measurements in the same manner as in Figures~3d-3f.  Namely, if the CNT alignment direction coincides with the groove direction, one should see maximum absorbance at $\theta$ = 0$^\circ$ and minimum absorbance at $\theta$ = 90$^\circ$.  Figure~5c shows that the absorbance becomes maximum at $\theta$ = 0$^\circ$ and minimum at $\theta$ = 90$^\circ$ for all spots examined, demonstrating that the CNTs globally aligned along the intentionally created groove direction.


In this study, we examined the question of what determines the CNT alignment direction in the global alignment process that occurs in the controlled vacuum filtration method.  Through a series of systematic experiments, we obtained unambiguous evidence that the direction of parallel grooves pre-existing on the surface of the filter membrane dictates the CNT alignment direction.  Furthermore, we showed that global alignment does not occur if the pre-existing grooves are eliminated from the membrane.  Finally, after eliminating pre-existing grooves, we intentionally created periodically spaced parallel grooves on the membrane using a diffraction grating.  The created grooves successfully defined the direction of the global alignment of CNTs in a precise and reproducible manner. 

\section{Methods} 
\subsection*{Suspension preparation} 
We purchased arc discharge SWCNTs (P2-SWNT) with an average diameter of 1.4~nm from Carbon Solution, Inc. Using sodium deoxycholate (DOC), purchased from Sigma-Aldrich, as surfactant, we suspended P2-SWNTs with an initial concentration of 0.4~mg/mL in DOC (0.5\% (wt./vol.)) by bath sonication for 15 minutes. We then further sonicated the obtained suspension with a tip sonicator (XL-2000 Sonicator from Qsonica, LLC.\ 1/4 inch probe) for 45 minutes at a power of 30~W. 
Next, we centrifuged the suspension for 1.5 hours at 38000~rpm (Beckman-Coulter Optima L-80 XP Preparatory Ultracentrifuge (BSI414) using a Beckman SW-41 Ti swing bucket rotor). After centrifugation, we collected the upper 80\% of the supernatant. Before the filtration process, we further diluted the suspension with pure water so that surfactant concentration be less than the critical micelle concentration.

\subsection*{Controlled vacuum filtration (CVF)} The filtration systems (Millipore\textsuperscript{\textregistered} XX1002500 glass microanalysis) used in the study was purchased from Fisher Scientific Company, LLC. It consisted of a 15-mL glass funnel, a clamp, a fritted glass filter support, and a silicone stopper. We purchased filter membranes with a pore size of 200~nm (Nuclepore Track-Etched Polycarbonate (Hydrophilic) Membranes, product number 110606 and 111106) from GE Healthcare Life Sciences.

We mounted each filtration setup on a side-arm flask, which was connected with a water flow vacuum pump (water aspirator). A filter membrane with a proper pore size was put on the surface of the glass support; liquid was poured into the glass funnel, which was sealed with the glass support by the clamp. A SWCNT suspension was poured into the glass funnel. At the beginning, we applied zero pressure to the filtration system. 
Near the end of the filtration process, the filtration rate was increased so as to stabilize the alignment structures and make the film uniform. This speedup process was started  before the concave meniscus of the remaining suspension touched the surface of the filter membrane. With an initial suspension volume of 6~mL, we started the acceleration process when there was 1~mL of suspension left. The pressure was controllable by adjusting multiple valves in the vacuum line; we monitored the value of the pressure by the pressure gauges in the vacuum line. After the completion of the filtration process, we kept the vacuum pump running for 15--30 minutes while the film was still on the membrane to dry the film. 

We used a wet transfer process to remove the filter membrane and place the SWCNT film on a desired substrate. First, we deposited a drop of pure water on the substrate and then placed the sample on the substrate with the side of the SWCNT film touching the substrate surface. Then we gently air-dried the film on the substrate so that the film stuck to the surface of the substrate firmly. We then immersed the substrate in chloroform for twenty minutes to dissolve the filter membrane. Finally, we cleaned the substrate by acetone, followed by pure water, and dried it by a gentle air flow.

When a filter membrane was cut into pieces before filtration, we placed them on the surface of the glass support by overlapping each piece (the width of overlapping area was less than 1~mm) such that they cover the whole glass support surface within the area of the glass funnel. We then followed the standard CVF procedure described above.

\subsection*{Characterization of alignment direction} We characterized the alignment direction of fabricated films by studying polarization-dependent absorbance using a 660-nm diode laser. The incident light beam was linearly polarized by a polarizer and then went through a half-wave plate, which was rotated by a rotation stage. The beam size at the sample was 1~mm. From the transmitted light intensity, we calculated the absorption value at each angle $\theta$, which is the angle between the light polarization and the specified axis on the film. Maximum absorption occurred when the polarization was parallel to the CNT alignment direction, while the amount of absorption was minimum when the polarization was perpendicular to the alignment direction. 

\subsection*{Characterization of surface morphology} Surface morphology of filter membranes was characterized by an optical microscope (Nikon, ECLIPSE LV100ND), an atomic force microscope (AFM) (Park Systems, Park NX20 for Figures 3 and 4, and BRUKER, MultiMode 8 for Figure 5), and a scanning electron microscope (SEM) (FEI, Quanta 400). The alignment structure (on micro- and nanometer scales) was examined by an AFM.

\subsection*{Eliminating grooves through heating treatment}  The filter membrane is made of polycarbonate, which has a glass transition temperature of 147\,$^\circ$C. Above this temperature, the polymer chains of polycarbonate become mobile.  Therefore, pressing a heated membrane against a flat surface can flatten the membrane surface. 
We thus heated the filter membrane to 163\,$^\circ$C and pressed it against a glass slide to flatten its surface using the following process. 
We put a glass slide on the hotplate, and then put a filter membrane with the smooth side down. We put another glass slide (perpendicular to the first one) on the filter membrane, and left them for three and half minutes. We then pushed the glass slide, applying force vertically to the filter membrane at multiple positions, for five and half minutes. 
We thoroughly checked multiple regions of a heat treated membrane by SEM and observed complete elimination of macrogrooves and microgrooves from the membrane surface.

\subsection*{Imprinting grooves using gratings} 
Gratings with groove densities of 300, 600, and 1800/mm (500~nm Blaze Wavelength Reflective Diffraction Gratings, product number GR25-0305, GR25-0605, and GR25-1850, respectively) were purchased from Thorlabs.
%
We put a grating with the groove side up on the hotplate, and then put a filter membrane with the smooth side down such that the smooth side of the filter membrane be in direct contact with the groove side of the grating. We put a glass slide on the filter membrane, and left them for three and half minutes. We then pushed the glass slide, applying force vertically to the filter membrane at multiple positions, for five and half minutes. 

\section{Supporting Information}
The Supporting Information is available free of charge.

Information on characteristics of filter membranes without any treatment, characteristics of filter membranes and single-wall carbon nanotube (SWCNT) films after heating treatment, and characteristics of filter membranes with intentionally created grooves and SWCNT films (PDF)

\begin{acknowledgement}

We acknowledge support by the Basic Energy Science (BES) program of the U.S.\ Department of Energy through Grant No.\ DE-FG02-06ER46308 (for preparation of aligned carbon nanotube films), the U.S.\ National Science Foundation through Grant No.\ ECCS-1708315 (for optical measurements), and the Robert A.\ Welch Foundation through Grant No.\ C-1509 (for structural characterization measurements).  The authors would like to acknowledge the staff and facilities of the Shared Equipment Authority at Rice University.  We also thank Timothy Noe, Ming Zheng, and Shigeo Maruyama for helpful discussions and Junko Eda and Hitomi Okubo for assistance in measurements.

\end{acknowledgement}






\providecommand{\latin}[1]{#1}
\providecommand*\mcitethebibliography{\thebibliography}
\csname @ifundefined\endcsname{endmcitethebibliography}
  {\let\endmcitethebibliography\endthebibliography}{}

\end{document}